\documentclass[%
prl, reprint%
 ,twocolumn%
 ,secnumarabic%
,amssymb, amsmath, nobibnotes, floatfix, citeautoscript,aps,superscriptaddress, showpacs]{revtex4-1}
\usepackage{bm}%
\usepackage[T1]{fontenc}
\expandafter\ifx\csname package@font\endcsname\relax\else
 \expandafter\expandafter
 \expandafter\usepackage
 \expandafter\expandafter
 \expandafter{\csname package@font\endcsname}%
\fi
\ifx\pdfoutput\undefined
\usepackage{graphicx}
\else
\usepackage[pdftex]{graphicx}
\usepackage{epstopdf}
\fi
\usepackage{hyperref}

\usepackage{natbib}
\bibpunct{ [}{]}{,}{n}{,}{,}

\begin{document}
\bibliographystyle{ieeetr}
\makeatletter 
\makeatother

\title{Dynamic parity recovery in a strongly driven Cooper-pair box}

\author{S. E. de Graaf}
\email{degraaf@chalmers.se}\affiliation{Department of Microtechnology and Nanoscience, MC2, Chalmers University of Technology, SE-41296 Goteborg, Sweden}
\author{J. Lepp$\rm{\ddot{a}}$kangas}\affiliation{Department of Microtechnology and Nanoscience, MC2, Chalmers University of Technology, SE-41296 Goteborg, Sweden}
\author{A. Adamyan}\affiliation{Department of Microtechnology and Nanoscience, MC2, Chalmers University of Technology, SE-41296 Goteborg, Sweden}
\author{A. V. Danilov}\affiliation{Department of Microtechnology and Nanoscience, MC2, Chalmers University of Technology, SE-41296 Goteborg, Sweden}
\author{T. Lindstr$\rm{\ddot{o}}$m}\affiliation{National Physical Laboratory, Hampton Road, Teddington, TW11 0LW, UK}
\author{M. Fogelstr$\rm{\ddot{o}}$m}\affiliation{Department of Microtechnology and Nanoscience, MC2, Chalmers University of Technology, SE-41296 Goteborg, Sweden}
\author{T. Bauch}\affiliation{Department of Microtechnology and Nanoscience, MC2, Chalmers University of Technology, SE-41296 Goteborg, Sweden}
\author{G. Johansson}\affiliation{Department of Microtechnology and Nanoscience, MC2, Chalmers University of Technology, SE-41296 Goteborg, Sweden}
\author{S. E. Kubatkin}\affiliation{Department of Microtechnology and Nanoscience, MC2, Chalmers University of Technology, SE-41296 Goteborg, Sweden}

\date{\today}
\begin{abstract}
We study a superconducting charge qubit coupled to an intensive
electromagnetic field and probe changes in the resonance frequency
of the formed dressed states. At large driving strengths, exceeding
the qubit energy-level splitting, this reveals the well known Landau-Zener-St$\rm{\ddot{u}}$ckelberg (LZS) interference structure of a longitudinally driven two-level system.
For even stronger drives we observe a significant change in the LZS pattern and contrast.
We attribute this to photon-assisted quasiparticle
tunneling in the qubit. This results in the recovery of the qubit parity, eliminating effects of quasiparticle poisoning and leads to an enhanced interferometric response.  
The interference pattern becomes robust to quasiparticle poisoning and has a good potential for accurate charge sensing.
\end{abstract}

\maketitle
In a two-level system (TLS) under strong periodic non-adiabatic driving, the relative phase accumulated between successive tunneling events can play a significant role. The constructive or destructive interference between consecutive tunneling events are generally referred to as  Landau-Zener-St$\ddot{\rm{u}}$ckelberg (LZS) oscillations.
LZS oscillations are a celebrated phenomenon observed in a variety of quantum systems, ranging from optical lattices\cite{kling2010}, nanomechanical circuits\cite{lahaye}, single spins in diamond NV-centers\cite{huang2011}, and semiconductor quantum dots\cite{petta,ribeiro1}, to Josephson qubits\cite{review,chris1,chris2,sill,oliver2005,oliver,izmalkov}.

In the regime of strong driving the steady state of a TLS coupled to a cavity is conveniently described in terms of photon dressed states, capturing the hybridization of the intense electromagnetic field and the TLS\cite{chris1}.
Such dressed states have successfully been applied in several applications, including lasing and amplification\cite{oelsner2013,astafiev2007}, suppression of decoherence in two-level systems \cite{timoney2011}, and single photon emission\cite{flagg2009}. 
It is especially attractive to study dressed states in superconducting microcircuits, not only due to
realizations of previously hardly accessible regimes\cite{chris1,chris2}, but also since they offer straightforward integration with other systems.



In this article we present experimental observations of the interplay between photon dressed states of a superconducting charge qubit (Cooper-pair box, CPB) and quasiparticle creation.
Remarkably, we find that when the CPB is driven into a regime where quasiparticle generation provides a dominant path for excitation/relaxation, the CPB exhibits a more pronounced interferometric response, while typically quasiparticle ''poisoning'' is degrading the performance of superconducting devices\cite{aumentado,lenander2011,catelani2011,juha2}.
This effect emerges for strong driving of the qubit, when the two-level approximation of the CPB breaks down, and it is observed as a significant distortion of the typical LZS interference pattern.
We attribute the increased response to the recovery of the qubit parity emerging in the ultra-strong driving regime.
The populations of dressed states changes due to new quasiparticle-induced transition channels. At sufficient drive strengths the observed non-linear process involves the coherent exchange of energy  equivalent of $\sim14$ photons between the cavity and the CPB. This coherence is broken by pair-breaking that quickly causes the qubit to relax to a specific charge state, recovering its parity.
The quasiparticle tunneling rates stay below the dressed state energy-level splittings, and do not destroy their coherence, but significantly influence their populations.

We find very good agreement with our simplified theory\cite{thypaper} despite the rather complex physics involved. 
Not only does our model provide a means for extended qubit characterization\cite{review}, but the observed effect also has high potential for very sensitive charge detection. The discussed regime can be used to force the CPB into a desired parity state, making charge detection robust with respect to quasiparticle poisoning and temperature.
\begin{figure}[t!]
\begin{centering}
\includegraphics[height=4.5 cm]{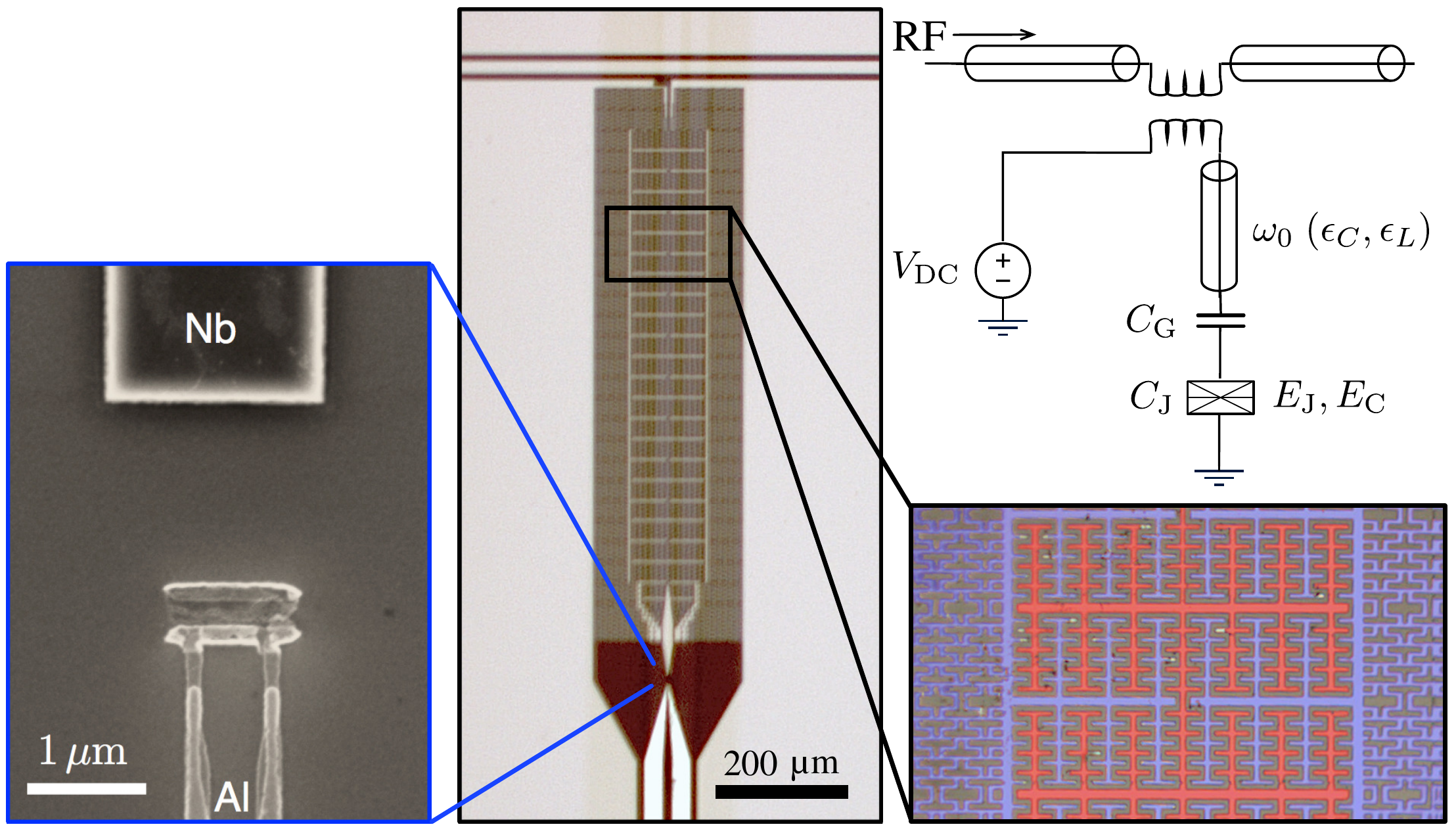}
\caption{Optical image and scanning electron micrograph of the sample. 
Bottom right: Colored close-up of a segment of the resonator (red) and nearby vortex-free ground plane (blue).   Top right: Circuit representation of the sample.}
\label{fig:sample}
\end{centering}
\end{figure}

\begin{figure*}[!]
\begin{centering}
\includegraphics[height=15.2 cm, angle=270,origin=b]{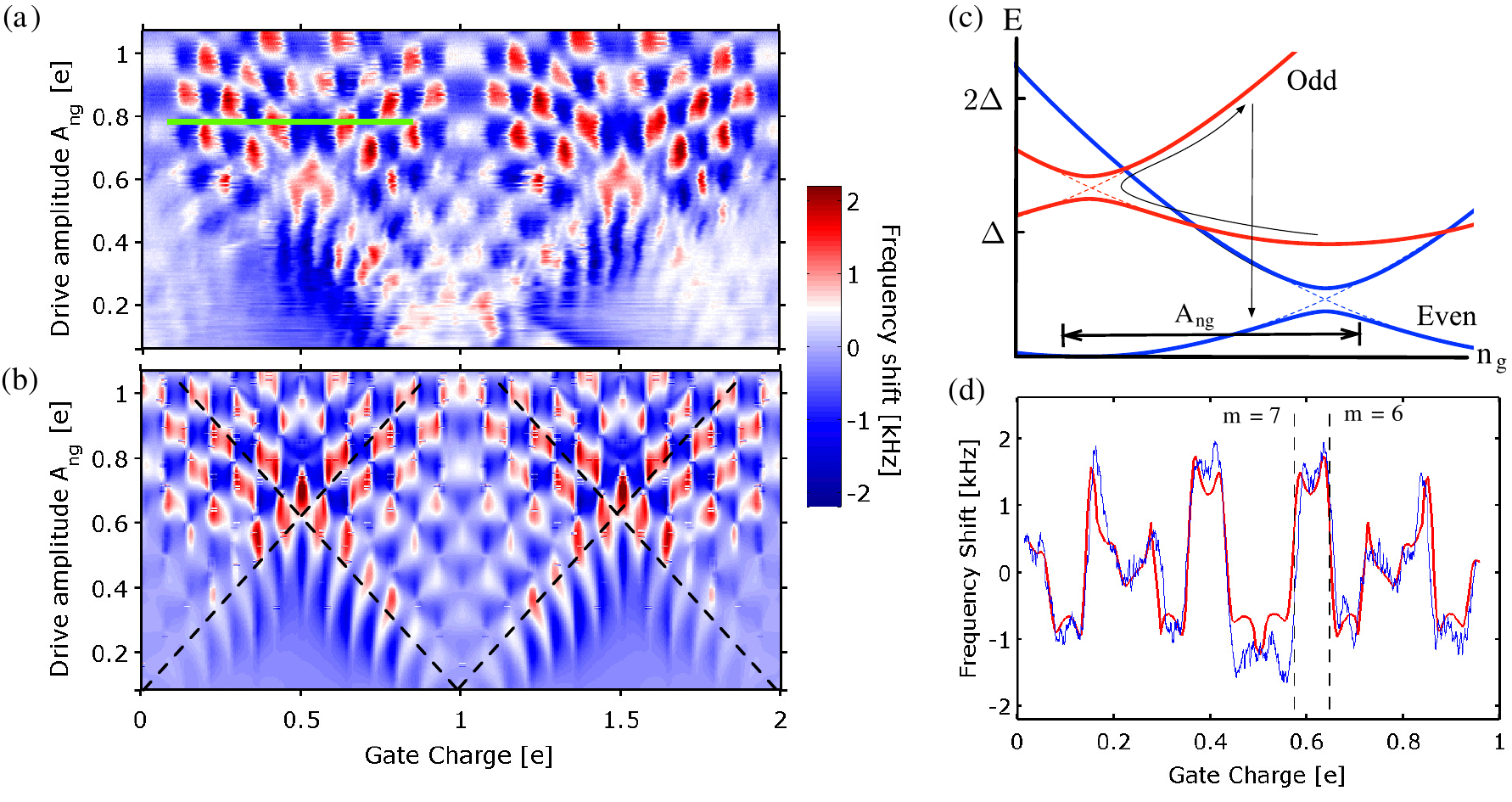}
\caption{(a) Measured frequency shift of the microwave cavity as a function of gate charge and microwave amplitude. The frequency shift indicates the time-averaged projections of the qubit state onto $\sigma_z$. T = 20 mK. (b) Numerical simulation. (c)  Simplified quasi-classical energy diagram in the charge basis illustrating the process of restoring the parity of the box. (d) Cross section of the raw data (blue), sampled at a rate of 50 kHz and averaged 8 times over a total time of 2.2 s per gate trace, and simulation (red) taken at the green line indicated in (a). For the best fit, a longitudinal Gaussian convolution with the width $\sigma_{n}= 0.0052$e is applied to the theoretical result, consistent with slow fluctuations of nearby charges at the given measurement timescale.
Dashed lines show the center of the 7th and 6th photon resonances.  In our sample we extract $C_G/C_J = 1/15$, $C_J=0.741$ fF, $\omega_0/2\pi=6.94$ GHz, $E_J= 4.82$ GHz, $E_C= 24.4$ GHz. The near-critically coupled cavity is excited with a microwave power of -90 to -100 dBm. To fit the experimental data we use an effective charge fluctuator and quasiparticle temperature $T_{CF}=100$ mK and $T_{qp}=200$ mK, a high frequency ohmic coupling constant\cite{leggett} $\alpha = 1.2\cdot 10^{-4}$, quasiparticle tunneling asymmetry $\Gamma_{\rm{odd}}=10^{7}\cdot\Theta(E_i-E_f)$ Hz and $G=0.1$.}
\label{fig:plot}
\end{centering}
\end{figure*}

Our sample, shown in Fig. \ref{fig:sample}, consists of an aluminum CPB capacitively coupled to a high quality ($Q\sim 40000$) niobium quarter-wave coplanar resonator on sapphire. We have designed the sample such that we apply microwave excitation and DC gating of the CPB via the same gate electrode, i.e. the center conductor of the resonator. We excite and read out the state of the box using the same cavity, which is inductively coupled to a transmission line in order to further reduce coupling to parasitic modes and effectively filter out the electromagnetic environment from the qubit\cite{kim2012,houck2008}.  The cavity is also designed to be free of Abrikosov vortices and trapped flux in both the resonator and the nearby ground, hence its rather unconventional layout (it is still conceptually equivalent to a $\lambda/4$ resonator)\cite{degraaf2012}. 

Not only do we need a high Q cavity to measure the weak dispersive shifts in the regime we are interested in, but to increase the measurement accuracy
we employ a technique called Pound-Drever-Hall (PDH) locking\cite{black,tobias}, commonly used in frequency metrology. A phase modulated (PM) signal is sent to the resonator, and when probed off resonance the PM sidebands interfere and cause an amplitude modulation (AM) of the carrier which can be easily detected with lock-in measurements. A feedback loop tries to minimize the AM signal, resulting in an analog signal that is directly proportional to the center frequency of the cavity.

The measurement of the CPB is presented in Fig. \ref{fig:plot}a and the numerical simulation in Fig. \ref{fig:plot}b. 
First, we will note the main features and give a very brief summary of their origin. 

(i)
For the microwave drive amplitude $A_{ng}\l\lesssim 0.5e$ we observe a typical LZS-type pattern as we slowly sweep the gate charge $n_G$.
The qubit relaxation drives the system to the ground state in the dressed basis,
except below the Coulomb-triangles indicated in Fig. \ref{fig:plot}b, where a bimodal structure
appears due to population inversion in the qubit basis. 
This bimodality has previously been explained in detail, and is known to vanish in the presence of too strong environmental charge relaxation and dephasing ($\lambda_{\rm{rel}}$)\cite{chris1, smirnov2003, tannoudji}. Non-equilibrium quasiparticles lead to 1e-periodicity which can be modeled by simply superimposing the same (but differently weighted) LZS-pattern with an offset of 1e.

(ii) In the region $A_{ng}\gtrsim0.5e$
the pattern shows almost rectangular features, with very sharp transitions from constructive to destructive interference along the gate axis. This bimodality is also a result of population inversion, driven by photon-assisted quasiparticle tunneling. Notable is that the contrast is strongly increased. A simplified quasi-classical picture is given in Fig. \ref{fig:plot}c. Driving the system near the even degeneracy produces coherent dynamics, however, when the CPB is poisoned by a quasiparticle, the coherence is lost (odd state) and the contrast is reduced. When the photon field $A_{ng}$ becomes large enough and the system enters the odd state it will quickly be driven above 2$\Delta$ in energy and a pair-breaking event quickly recovers the parity, and therefore also the contrast. {\it{The pattern becomes immune to nonequilibrium quasiparticle poisoning.}}

To continue with a detailed modeling of our system we start by observing that
the data at drives $A_{ng}\gtrsim0.5e$ cannot be explained simply by superimposing the structure of the intermediate-drive region with probabilities to be in the two electron-parity subspaces.
To understand processes behind the interference pattern at higher drive strengths, we model the coherent evolution of the coupled cavity-CPB system with the Hamiltonian\cite{juha1}
\begin{equation}\label{Hamiltonian1}
\begin{split}
 H_0 &= \hbar\omega_0 a^\dagger a + E_C(\hat{n}-n_G)^2 + \hat{n}g(a^\dagger + a)  \\
 &- \frac{E_J}{2}\sum_n\left(|n+2\rangle\langle n| + |n\rangle\langle n+2|\right),
\end{split}
\end{equation} 
with cavity frequency $\omega_0/2\pi$.
The island charge  $\hat n$, counted in single electrons,
may take odd integer values in the presence of quasiparticle tunneling. Furthermore, $E_C=e^2/2(C_J+C_G)$, and the term containing the Josephson energy, $E_{\rm J}$, constitutes 
coherent Cooper-pair tunneling.
The gate capacitor $C_{\rm G}$ couples the cavity and CPB bilinearly with coupling constant
$g=\hbar\omega_0 G \times C_G/C_J$. Here $G=(2\epsilon_C/\epsilon_L)^{1/4}$ with $\epsilon_L$ and $\epsilon_C$ the magnetic and charging energy of the cavity.

A common basis for further analysis is the
displaced number states of (\ref{Hamiltonian1}) for $E_{ J}=0$\cite{tannoudji}, 
$
\vert n;N\rangle^{0}=  \exp[-\hat{n}\beta( a^{\dagger}- a)]\vert n\rangle\vert N\rangle,
$ with eigenenergies $E_{n;N}^{0}=N\hbar\omega_0-n^2\frac{ g ^2}{\hbar\omega_0} +E_C(n-n_G)^2.$
Here $\vert n\rangle$ corresponds to the fixed island charge state and $\vert N \rangle$ to the photon (Fock) state, and $\beta=g/\hbar\omega_0$. 
For $E_{J}\neq 0$ Cooper-pair tunneling mixes these states with amplitudes
$
E_J\langle n\pm 2;N+m\vert n;N\rangle\approx E_JJ_{m}(\pm a_N)\equiv E_J^{\rm{dr}},
$ (in the limit $N\gg 1$)
where $J_m$ are Bessel functions, and $a_N\equiv 4\beta\sqrt{N} = 4E_CA_{\rm{ng}}/\hbar\omega_0$.
The dressed two-level system is then given in the charge basis as 
\begin{equation}
\begin{matrix}\vert -,N\rangle_{e/o}=\cos\phi\vert n,N\rangle+\sin\phi\vert n+2,N-m\rangle\hspace{0.9mm} \\ \vert +,N\rangle_{e/o}=\sin\phi\vert n,N\rangle-\cos\phi\vert n+2,N-m\rangle,\end{matrix}
\end{equation}
with $2\phi = \arctan{[E_{J}^{\rm{dr}}/(\delta E_C-m\hbar\omega_0)]}+\pi\Theta(m\hbar\omega-\delta E_C)$ and $\delta E_C$ being the charging energy difference. We define the even (e) and odd (o) parities depending on $n$ which, so far, are completely decoupled.

This hybridized two-level state describes the dressed equivalent of LZS oscillations in the absence of dissipation. It has
previously been studied using an additional adiabatic low frequency readout\cite{sill,chris1,chris2, izmalkov}. Here we demonstrate a direct readout via
the related shift in the resonant frequency of the photon-dressed CPB.

The populations of the dressed  states also affect the interference
pattern and are influenced mainly by two decoherence mechanisms.
First, gate charge fluctuations causes energy relaxation within fixed parity of the island ($\lambda_{\rm{rel}}$). We model this as an ohmic bath: $H_B = \hat n\sum_j\lambda_j(b_j^\dagger + b_j) + \sum_j\omega_jb_j^\dagger b_j$\cite{chris1, leggett}.
Second, not all Cooper-pairs stay paired in this nonequilibrium system, and subsequent quasiparticle tunneling can cause inter-parity transitions.
We perturbatively introduce incoherent transition rates between even and odd charge eigenstates of the Hamiltonian in our density-matrix simulation \cite{thypaper}
\begin{equation}
\begin{split}
\Gamma_{i\rightarrow f} &=  \bigg[\Gamma_{\rm{odd}}+\frac{1}{R_Te^2}\int \eta(\omega)\left[1-\eta(\omega+\delta E)\right]\rho(\omega)   \\
&\times \rho(\omega+\delta E)d\omega\bigg](|\langle f |\hat{T}|i\rangle|^2 + |\langle f |\hat{T}^{\dagger}|i\rangle|^2).
\end{split}\label{eq:rate}
\end{equation}
Here $\eta(\omega)$ is the general quasiparticle distribution, $\rho=\Theta(\omega^2-\Delta^2)\omega/\sqrt{\omega^2-\Delta^2}$ is the BCS quasiparticle density of states, $\Delta$ the superconductor energy gap, and $\Gamma_{\rm{odd}}$ accounts for the asymmetry of having an extra quasiparticle on the island\cite{schwind}. $\delta E$ is the dressed state
energy difference between the two states $\vert i\rangle$ and $\vert f\rangle$, and $\hat{T}=\sum_n|n+1\rangle\langle n|$ is the corresponding operator for single electron tunneling. It results in matrix elements of the form $\langle n\pm 1;N+m\vert\hat{T}+\hat{T}^\dagger\vert n;N\rangle\approx J_m(\pm a_N/2)$, with twice the period of the Cooper pair elements.
Dynamics of the created quasiparticles at stronger drive can be neglected,
as in this regime the results are very robust with respect to changes in the  quasiparticle distribution $\eta$.
We numerically
diagonalize (\ref{Hamiltonian1}) near selected values of $N\gg 1$ with the assumption of a locally constant $J_m(a_N)$.
The eigenstates can then be grouped into "equivalent" states, which differ only by the energy of a photon translation (Wannier-Stark ladder). 

After we have obtained the populations for the dressed states, the frequency shift can be approximated from the small shifts in the energy-level structure when moving in the photon ladder.
At resonance, the frequency shift of the cavity for single photon jumps in the dressed ladder will have a magnitude $\Delta E= (E_{N+1}-E_N)-\hbar\omega_0 \propto \pm [ J_m(a_{N+1})- J_m(a_N) ]$.
Linearizing the Bessel functions around $N$, the measured frequency of the cavity for a given photon number and charge state becomes
\begin{equation}
\hbar \omega_\pm\approx \hbar\omega_0\pm\bigg[J_{m-1}(a_N) -J_{m+1}(a_N) \bigg]\frac{E_JG^2C_G^2}{2 a_NC_J^2},
\label{eq:coupling}
\end{equation} 
where $\pm$ indicates the charge state in the even parity.

\begin{figure}[t!]
\begin{centering}
\includegraphics[height=5.8cm]{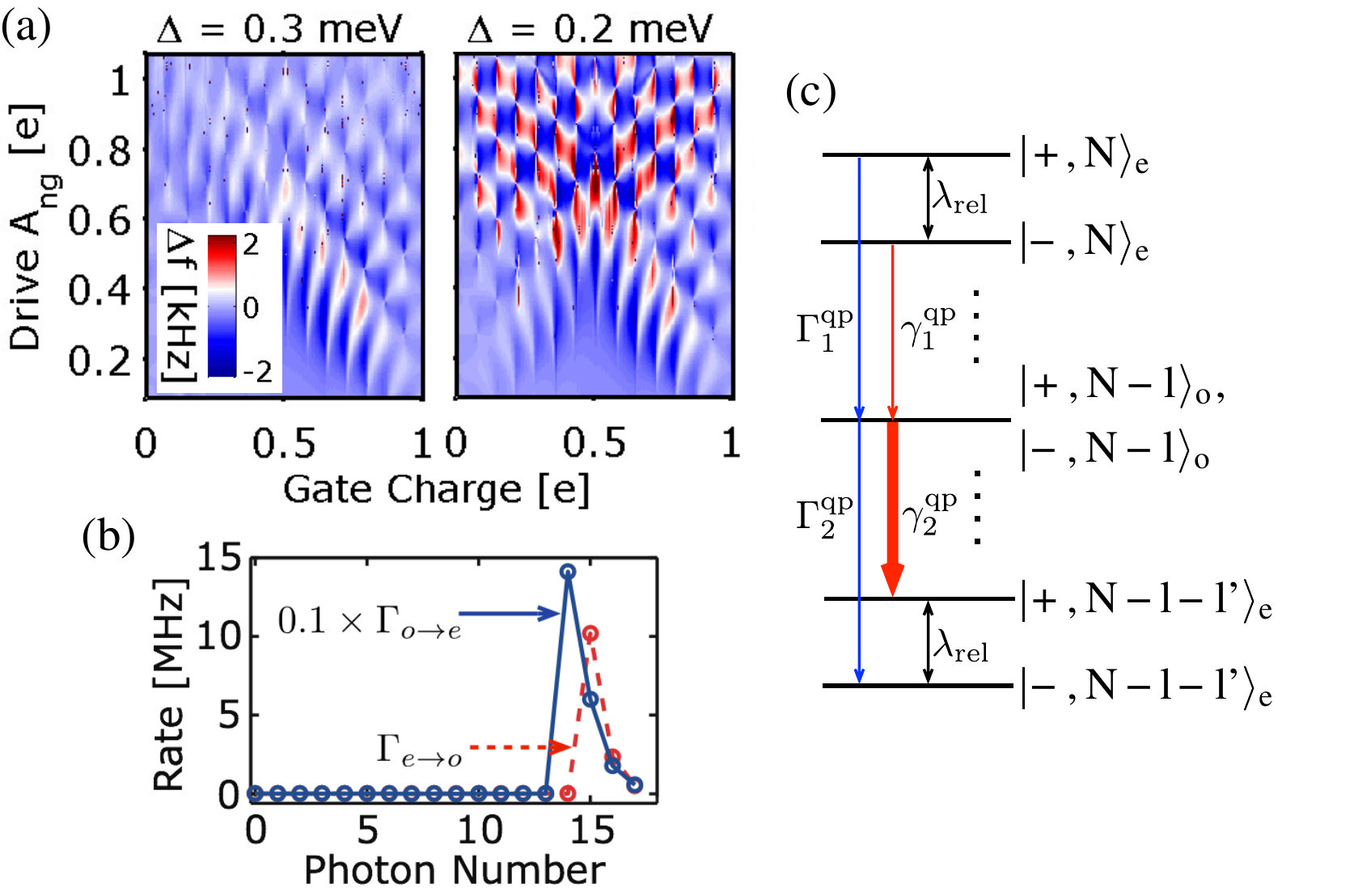}
\caption{(a) Calculated frequency shift as a function of photon number and charge offset for the coupled cavity-CPB for two values of the superconducting gap $\Delta$. (b) Calculated transition rate between dressed states of the two subspaces as a function of number of dissipated photons evaluated for $A_{\rm{ng}}=0.9$ and $m=4$. The two directions (note the different scaling) are shifted by one photon due to small charging energy differences. (c) Energy level representation for the photon-assisted relaxation via the quasiparticle subspace that results in population inversion (when $\gamma_2^{\rm{qp}}>\Gamma_2^{\rm{qp}}$) and fast recovery of the even parity. }
\label{fig:sim}
\end{centering}
\end{figure}

Our numerical simulation for the frequency shift is compared to the experimental data in Fig. \ref{fig:plot}.
We find very good agreement for a wide range of drive powers. For low drives significant quasiparticle tunneling leads to a reduced contrast, since it switches the system between the two subspaces slowly in a probabilistic manner. The intrinsic relaxation due to charge fluctuators  ($\lambda_{\rm{rel}}$) are responsible for the 'droplet'-shaped interference fringes observed. In a system without quasiparticles these would be present everywhere inside the interference region (see Fig. \ref{fig:sim}a).

For strong drives a different type of process sets in and dominates the dressed state transitions: multi-photon absorption with energy $>2\Delta$. The energy is used to generate a single-electron tunneling event across the junction, changing the parity of the island.
Each quasiparticle tunneling event occurs through the absorption of  $l\hbar\omega_0\approx 2\Delta=14$ photons, as shown in Fig. \ref{fig:sim}b. The system can make a transition, for example, from $\vert +,N\rangle_e$ to the state $|+,N-l\rangle_o$ with rate $\Gamma^{\rm{qp}}_1$, as illustrated in Fig. \ref{fig:sim}c. We define two different rates, $\Gamma^{\rm{qp}}_1$ and $\gamma^{\rm{qp}}_1$ (both $\approx \Delta |J_l(a_N/2)|^2/e^2R_T$ with different $l$ due to charging energy differences), that distinguish states within the dressed manifold for the non-linear dissipation of the coherent process that exchanges the energy of $\sim14$ photons between cavity and CPB. These rates differ depending on bias point; however, since they immediately result in the loss of coherence, their relative magnitudes do not matter; they bring the system to the (effectively) same state. Only the subsequent processes have a significant impact on the qubit populations.

\begin{figure}[b!]
\begin{centering}
\includegraphics[height=3.55cm]{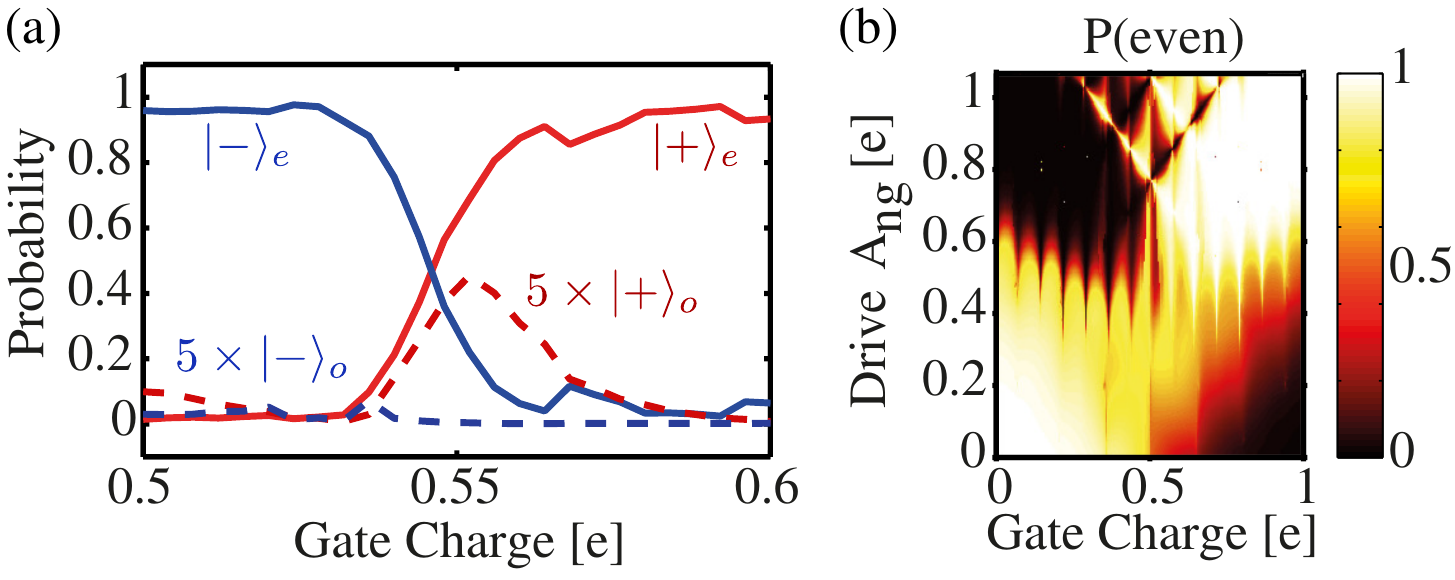}
\caption{(a) Calculated occupation probability for the even (solid line) and odd (dashed line) dressed states at the $m=7$ even photon resonance for $A_{ng}=0.65$.  (b) Calculated occupation probability of the even parity state (summed over all even charge states). At low drive the incorrect parity (the parity that does not result in coherent dynamics) is dominating, while the proper parity is almost completely recovered at strong drives.}
\label{fig:par}
\end{centering}
\end{figure}

Once in the odd subspace the system can create another pair of quasiparticles, now substantially faster, with the help of an increased charging energy through coherent Cooper-pair tunneling within $|+,N-l\rangle_o$ or $|-,N-l\rangle_o$. Through this energetic charge state the system quickly returns to the original parity since most of the required energy $\sim 14\hbar\omega_0$ is already stored as charging energy and the additional number of photons required, $\sim (2\Delta-\delta E_C)/\hbar\omega_0$, is much less. Now relaxation is preferred into one of the states $|+,N-l-l'\rangle_e$ or $|-,N-l-l'\rangle_e$, as illustrated by $\Gamma^{\rm{qp}}_2$ and $\gamma^{\rm{qp}}_2$ ($\approx \Delta |J_{l'}(a_N/2)|^2/e^2R_T$) in Fig. \ref{fig:sim}c. The rate that results in the eigenstate with the lowest charging energy will be dominating, and this results in a sudden change in population on one side of the photon resonance, see Fig. \ref{fig:par}a. Exactly at the $m$-photon resonance there is an enhanced occupation of the odd state owing to increased quasiparticle generation. Due to the almost instant relaxation from the  odd high energy state its occupation probability is essentially zero. This can also be understood from Fig. \ref{fig:sim}b, where the escape rate from the odd subspace is roughly 10 times faster than the rate that populates it, and this slow rate sets an upper limit for non-equilibrium quasiparticle tunneling. This type of population inversion resembles the one previously studied at lower drive strengths\cite{chris1}, and depends strongly on the relaxation to the bath. 
Similar relaxation mechanisms have also been observed in single artificial atom lasing experiments\cite{astafiev2007} and can be related to the Josephson quasiparticle (JQP) cycle.

At low drives the system is poisoned by quasiparticles, and most of the time is spent in the parity that does not give any coherent dynamics, see Fig. \ref{fig:par}b. When 
photon-assisted quasiparticle generation starts to dominate over nonequilibrium quasiparticle tunneling the system recovers its parity, hence the increase in contrast. In this regime $\Gamma_{2}^{qp}$ becomes very fast, since the $|-1e\rangle$ state becomes accessible, constantly "resetting" the qubit into the even parity (Fig. \ref{fig:plot}c). 

To show the implications of the superconducting gap on quasiparticle generation we increase and decrease the value of $\Delta$ in the simulation, see Fig. \ref{fig:sim}a. The best agreement corresponds to a typical aluminum thin film energy gap $\Delta\approx 0.2$ meV (49 GHz). For a larger gap there is no quasiparticle generation taking place and no enhanced contrast or bimodality.

As  suggested by Sillanp$\rm{\ddot{a}\ddot{a}}$ et al. \cite{sill} this kind of device could have an application as a very good charge sensor due to its many oscillations within one charge period. For this purpose the quasiparticle induced bimodal structure shows even better potential as a charge sensor due to the sudden population inversion along the gate axis. 
From the data in Fig. \ref{fig:plot}d we extract a sensitivity of $2.9\pm0.6$ $\rm{\mu e}/\sqrt{\rm{Hz}}$ in the region $0.5 <A_{ng} <1$, comparable to state of the art single-electron transistors\cite{chargesensrecord}. This despite the fact that the system was not initially designed having charge sensitivity in mind, and the measured dispersive shifts are $\sim1\%$ of the cavity linewidth. Increasing the ratio $C_G/C_J$ a few times would not significantly alter the interplay between different relaxation rates, but it would increase the measured frequency shifts by a large amount (cf. eq. \ref{eq:coupling}). Furthermore, in the strong driving regime such a charge sensor would be unaffected by thermal quasiparticle poisoning since the odd and even states approach a steady-state population.

We aknowledge the Swedish Research Council VR, EU FP7 programme under the grant agreement 'ELFOS', the Marie Curie Initial Training Action (ITN) Q-NET 264034, and the Linnaeus centre for quantum engineering for financial support. 

\bibliographystyle{plainnat}

\begin{thebibliography}{26}
\footnotesize
\bibitem{kling2010} S. Kling, T. Salger, C. Grossert, and M. Weitz, \href{http://dx.doi.org/10.1103/PhysRevLett.105.215301}{Phys. Rev. Lett.,} {\bf{105}}, 215301 (2010).

\bibitem{lahaye} M. D. LaHaye, J. Suh, P. M. Echternach, K. C. Schwab, and M. L. Roukes, \href{http://dx.doi.org/10.1038/nature08093}{Nature}, {\bf{459}}, 960 (2009).

\bibitem{huang2011} P. Huang, J. Zhou, F. Fang, X. Kong, X. Xu, C. Ju, and J. Du, \href{http://dx.doi.org/10.1103/PhysRevX.1.011003}{Phys. Rev. X,} {\bf{1}}, 011003 (2011).

\bibitem{petta} J. R. Petta, H. Lu, and A. C. Gossard, \href{http://dx.doi.org/10.1126/science.1183628}{Science,} {\bf{327}}, 5966 (2010).

\bibitem{ribeiro1} H. Ribeiro, and G. Burkard, \href{http://dx.doi.org/10.1103/PhysRevLett.102.216802}{Phys. Rev. Lett.,} {\bf{102}}, 216802 (2009).



\bibitem{oliver2005} W. D. Oliver, Y. Yu, J. C. Lee, K. K. Berggren, L. S. Levitov, and T. P. Orlando
\href{http://dx.doi.org/10.1126/science.1119678}{Science}, {\bf{310}}, 1653 (2005).

\bibitem{sill} M. Sillanp$\rm{\ddot{a}\ddot{a}}$, T. Lehtinen, A. Paila, Y. Makhlin, and P. Hakonen, 
\href{http://dx.doi.org/10.1103/PhysRevLett.96.187002}{Phys. Rev. Lett.}, {\bf{96}}, 187002 (2006).
\bibitem{chris2} C. M. Wilson, T. Duty, F. Persson, M. Sandberg, G. Johansson, and P. Delsing,  \href{http://dx.doi.org/10.1103/PhysRevLett.98.257003}{ Phys. Rev. Lett.}, {\bf{98}}, 257003 (2007).

\bibitem{oliver} D. M. Berns, M. S. Rudner, S. O. Valenzuela, K. K. Berggren, W. D. Oliver, L. S. Levitov, and T. P. Orlando, \href{http://dx.doi.org/10.1038/nature07262}{Nature}, {\bf{455}}, 51 (2008).

\bibitem{izmalkov} A. Izmalkov, S. H. W. van der Ploeg, S. N. Shevchenko, M. Grajcar, E. Il'ichev, U. H$\rm{\ddot{u}}$bner, A. N. Omelyanchouk, H.-G. Meyer, \href{http://dx.doi.org/10.1103/PhysRevLett.101.017003}{Phys. Rev. Lett.}, {\bf{101}}, 017003 (2008).

\bibitem{review} S. N. Shevchenko, S. Ashhab, and F. Nori, \href{http://dx.doi.org/10.1016/j.physrep.2010.03.002}{Phys. Reports}, {\bf{492}}, 1 (2010).


\bibitem{chris1} C. M. Wilson, G. Johansson, T. Duty, F. Persson, M. Sandberg, and P. Delsing, \href{http://dx.doi.org/10.1103/PhysRevB.81.024520}{Phys. Rev. B}, {\bf{81}}, 024520 (2010).

\bibitem{astafiev2007} O. Astafiev, K. Inomata, A. O. Niskanen, T. Yamamoto, Yu. A. Pashkin, Y. Nakamura, and J. S. Tsai, \href{http://dx.doi.org/10.1038/nature06141}{Nature}, {\bf{449}}, 588 (2007).


\bibitem{oelsner2013} G. Oelsner, P. Macha, O. V. Astafiev, E. Il'ichev, M. Grajcar, U. H\"ubner, B. I. Ivanov, P. Neilinger, and H.-G. Meyer, \href{http://dx.doi.org/10.1103/PhysRevLett.110.053602}{Phys. Rev. Lett.}, {\bf{110}}, 053602 (2013).





\bibitem{timoney2011} N. Timoney, I. Baumgart, M. Johanning, A. F. Var\'on, M. B. Plenio, A. Retzker, and Ch. Wunderlich, \href{http://dx.doi.org/10.1038/nature10319}{Nature}, {\bf{476}}, 185 (2011).

\bibitem{flagg2009}E. B. Flagg, A. Muller, J. W. Robertson, S. Founta, D. G. Deppe, M. Xiao, W. Ma, G. J. Salamo, and C. K. Shih, \href{http://dx.doi.org/10.1038/nphys1184}{Nature Physics}, {\bf{5}}, 203 (2009).


\bibitem{aumentado} J. Aumentado, M. W. Keller, J. M. Martinis, and M. H. Devoret, \href{http://dx.doi.org/10.1103/PhysRevLett.92.066802}{Phys. Rev. Lett.}, {\bf{92}}, 066802 (2004).
\bibitem{lenander2011} M. Lenander, H. Wang, R. C. Bialczak, E. Lucero, M. Mariantoni, M. Neeley, A. D. O'Connell, D. Sank, M. Weides, J. Wenner, T. Yamamoto, Y. Yin, J. Zhao, A. N. Cleland, and J. M. Martinis, \href{http://dx.doi.org/10.1103/PhysRevB.84.024501}{Phys. Rev. B}, {\bf{84}}, 024501 (2011).

\bibitem{catelani2011} G. Catelani, J. Koch, L. Frunzio, R. J. Schoelkopf, M. H. Devoret, and L. I. Glazman, \href{http://dx.doi.org/10.1103/PhysRevLett.106.077002}{Phys. Rev. Lett.}, {\bf{106}}, 077002 (2011).

\bibitem{juha2} J. Lepp$\ddot{\rm{a}}$kangas, and M. Marthaler, \href{http://dx.doi.org/10.1103/PhysRevB.85.144503}{Phys. Rev. B}, {\bf{85}}, 144503 (2012).
\bibitem{thypaper} J. Lepp$\rm{\ddot{a}}$kangas, S. E. de Graaf, A. Adamyan, M. Fogelstr$\rm{\ddot{o}}$m, A. V. Danilov, T. Lindstr$\rm{\ddot{o}}$m, S. E. Kubatkin, and G. Johansson, \href{http://arxiv.org/abs/1306.4200}{arXiv:1306.4200} [cond-mat.mes-hall] (2013).

\bibitem{houck2008} A. A. Houck, J. A. Schreier, B. R. Johnson, J. M. Chow, J. Koch, J. M. Gambetta, D. I. Schuster, L. Frunzio, M. H. Devoret, S. M. Girvin, and R. J. Schoelkopf, \href{http://dx.doi.org/10.1103/PhysRevLett.101.080502}{Phys. Rev. Lett.}, {\bf{101}}, 080502 (2008).
\bibitem{kim2012}Z. Kim, B. Suri, V. Zaretskey, S. Novikov, K. D. Osborn, A. Mizel, F. C. Wellstood, and B. S. Palmer, \href{http://dx.doi.org/10.1103/PhysRevLett.106.120501}{Phys. Rev. Lett.}, {\bf{106}}, 120501 (2012).

\bibitem{degraaf2012} S. E. de Graaf, A. V. Danilov, A. Adamyan, T. Bauch, and S. E. Kubatkin, \href{http://dx.doi.org/10.1063/1.4769208}{J. Appl. Phys.},  {\bf{112}}, 123905 (2012).

\bibitem{black} E. D. Black, \href{http://dx.doi.org/10.1119/1.1286663}{Am. J. Phys.}, {\bf{69}}, 79 (2001).
\bibitem{tobias} T. Lindstr$\rm{\ddot{o}}$m, J. Burnett, M. Oxborrow and A. Ya. Tzalenchuk, \href{http://dx.doi.org/10.1063/1.3648134}{Rev. Sci. Inst.}, {\bf{82}}, 104706 (2011).


\bibitem{tannoudji} C. Cohen-Tannoudji, J. Dupont-Roc, and G. Grynberg, {\it{Atom-Photon Interactions}} (John Wiley, New York, 1992).
\bibitem{smirnov2003} A. Yu. Smirnov, \href{http://dx.doi.org/10.1103/PhysRevB.67.155104}{Phys. Rev. B}, {\bf{67}}, 155104 (2003).

\bibitem{juha1} M. Marthaler, J. Lepp\"akangas, and J. H. Cole, \href{http://dx.doi.org/10.1103/PhysRevB.83.180505}{Phys. Rev. B}, {\bf{83}}, 180505(R) (2011).

\bibitem{leggett} A. J. Leggett, S. Chakravarty, A. T. Dorsey, M. P. A. Fisher, A. Garg, and W. Zwerger, \href{http://dx.doi.org/10.1103/RevModPhys.59.1}{Rev. Mod. Phys.}, {\bf{59}}, 1 (1987).


\bibitem{schwind} G. Sch\"on, and A. D. Zaikin, \href{http://dx.doi.org/10.1209/0295-5075/26/9/010}{Eur. Phys. Lett.}, {\bf{26}}, 695 (1994).



\bibitem{chargesensrecord} H. Brenning, S. Kafanov, T. Duty, S. Kubatkin, and P. Delsing, \href{http://dx.doi.org/10.1063/1.2388134}{J. Appl. Phys.}, {\bf{100}}, 114321 (2006).

\end{thebibliography}

\end{document}